\documentclass[aps,prd,preprint,nofootinbib,superscriptaddress]{revtex4}

\usepackage{graphicx,graphics,epsfig}
\usepackage{dcolumn}
\usepackage{feynmf}
\usepackage{verbatim}
\usepackage{subfigure}
\usepackage{amsmath,amssymb,amsbsy,multirow,wasysym,bm,bbm}
\usepackage{slashed}

\usepackage{array,booktabs,colortbl,colordvi,multirow}
\usepackage{colordvi,color,xcolor}

\usepackage{rotating}

\newcommand{\newc}{\newcommand}
\newc{\bsym}{\boldsymbol}
\newc{\mrm}{\mathrm}
\newc{\ovl}{\overline}
\newc{\ovla}{\overleftarrow}
\newc{\ovra}{\overrightarrow}
\newc{\wtil}{\widetilde}

\newcommand{\Tr}{\mathrm{Tr}}

\newc{\eeq}{\end{equation}}
\newc{\beq}{\begin{equation}}
\newc{\ec}{\end{center}}
\newc{\bc}{\begin{center}}
\newc{\eeqa}{\end{eqnarray}}
\newc{\beqa}{\begin{eqnarray}}
\newc{\Beeqa}{\end{Beqnarray}}
\newc{\Bbeqa}{\begin{Beqnarray}}
\newc{\eref}[1]{Eq.(\ref{#1})}
\newc{\fref}[1]{FIG. \ref{#1}}
\newc{\sigv}{\left< \sigma v\right>}
\newc{\nuless}{$0\nu\beta\beta$}
\newc{\clr}{\color{red}}
\newc{\LH}{\hat{L}}
\newc{\RH}{\hat{R}}
\newc{\sWsq}{\sin^2\theta_\mathrm{W}}
\newc{\cWsq}{\cos^2\theta_\mathrm{W}}
\newc{\half}{\frac{1}{2}}
\begin{document}
\title{Electroweak Vacuum Stability and the Seesaw Mechanism Revisited}

\author{J.~N.~Ng}
\email[]{misery@triumf.ca}
\affiliation{Theory Department, TRIUMF, 4004 Wesbrook Mall, Vancouver, B.C., Canada }
\author{Alejandro de la Puente}
\email[]{apuente@physics.carleton.ca}
\affiliation{Department of Physics, Carleton University, 1125 Colonel By Drive, Ottawa, ON K1S 5B6 Canada }
\affiliation{Theory Department, TRIUMF, 4004 Wesbrook Mall, Vancouver, B.C., Canada }

\date{\today}

\begin{abstract}
We study the electroweak vacuum stability in Type I seesaw models for 3 generations of neutrinos in scenarios where the right-handed neutrinos have explicit bare mass terms in the Lagrangian and where these are dynamically generated through the mechanism of spontaneous symmetry breaking. To
best highlight the difference of the two cases we concentrate on the absolute stability of the scalar potential. We observe that for the first scenario, the scale at which the scalar potential becomes unstable is lower from that within the Standard Model. In addition the Yukawa couplings $\mathbf{Y}_\nu$ are constrained such that $\Tr{[\mathbf{Y}^{\dagger}_\nu \mathbf{Y}_{\nu}}] \lesssim10^{-3}$. In the second scenario the electroweak stability can be improved in a large region of parameter space. However, we found that the scalar used to break the lepton number symmetry cannot be too light and have a large coupling to right-handed neutrinos in order for the seesaw mechanism to be a valid mechanism for neutrino mass generation. In this case we have
$\Tr [\mathbf{Y}^\dagger_{\nu}  \mathbf{Y}_\nu]\lesssim 0.01$.
\end{abstract}

\maketitle

\section{Introduction}

The discovery of the Higgs boson at the LHC completes the Standard Model (SM) of particle physics. The measured value of the Higgs boson mass, $m_{h} = 125.09 \pm 0.21_{stat.}\pm 0.11_{sys.}$ as given by the combined results of both ATLAS and CMS experiments~\cite{ATLASCMS,ATLAS,CMS} implies that the quartic coupling, $\lambda_H$, of the SM scalar potential,  $V(H)$, is relatively small. Specifically, the scalar potential is defined by $V(H)=\frac{\lambda_H}{2}\left(H^\dagger H -v^2/2\right)^2$ where $H$ is the SM doublet Higgs field and $v=246$ GeV. When it is used for the boundary value of the renormalization group (RG) running of $\lambda_H$, it results in a negative value of $\lambda_{H}$ below the Planck scale due to the large top quark Yukawa coupling \cite{Bezrukov:2012sa,StrumiaNNLO,Alekhin:2012py,Buttazzo:2013uya}. In fact, the latest NNLO study on vacuum stability requires that $m_{h}>129.4\pm1.8$ GeV~\cite{StrumiaNNLO} for absolute stability up to the Planck scale. As a result, the electroweak vacuum is unstable at high energies. However, the most up-to-date measurements of the top quark mass and the strong coupling constant result in a scalar potential that is metastable with a sufficiently long-lived electroweak vacuum~\cite{Isidori:2001bm,Ellis:2009tp}. The exact value of the scale, which we denote by $\Lambda^{\text{SM}}_{I}$ where $\lambda_H$ becomes negative strongly depends on the value of the top quark mass and the strong coupling constant and it is well known that at the 2-loop level in the renormalization group equation evolution of SM couplings the value of $\lambda_{H}$ can turn negative at scales in the range $\sim 10^{10}-10^{13}$ GeV.  These remarks hold if we assume that no new physics enters in extrapolating the SM to such high energies.

Below we study the electroweak vacuum stability in Type I seesaw models for 3 generations of neutrinos in scenarios where the right-handed neutrinos have explicit bare mass terms in the Lagrangian and where these are dynamically generated through the mechanism of spontaneous symmetry breaking. We compare the two scenarios for the case of absolute stability of the scalar potential.

\section{Stability in the Type I Seesaw}
Despite its spectacular success the SM cannot be the complete theory of nature. We now have convincing evidence that neutrinos have small but finite masses. Within the SM this can arise by incorporating a dimension five effective operator $\kappa LLHH$, where $L$ denotes a SM lepton doublet. After electroweak symmetry breaking neutrinos obtain a mass given by $\frac{\kappa v^2}{2}$. By dimensional arguments, $\kappa$ scales as $1/M$ where $M$ is the new physics scale where the above operator is generated. The seesaw mechanism is the simplest manifestation of the above idea. One adds two or more right-handed neutrinos, $N_i$, where $i>2$ to the SM. In most scenarios, $i$ is taken to be 3 for symmetry reasons and for concreteness this is what we assume. In this simplest version the Lagrangian that gives rise to neutrino masses is given by
\beq
\label{eq:Lnu}
{\mathcal{L}}_\nu= -\ovl{N_R}{\mathcal{M}}_D\nu_L -\half\ovl{N_R} {\mathcal{M}}_M N_R^{c} + h.c.
\eeq
where ${\mathcal{M}}_D=\frac{1}{\sqrt 2}v {\mathbf {Y}}_\nu$ and ${\mathbf{Y}_\nu}$ is the $3\times3$ neutrino Yukawa coupling matrix. ${\mathcal{M}}_M$ is the Majorana mass matrix for the right-handed $N_i$ fields. Without lost of generality we shall take ${\mathcal{M}}_{M}$ to be diagonal and we will work in the charged lepton mass basis. It is well known that for $M>>v$ the masses of the light SM neutrinos are given by $m_\nu \simeq \frac{(y_{\nu}v)^2}{2M}$ as the result of integrating out the heavy right-handed neutrinos \cite{seesaw}. For energies below $M$, $\kappa$ is a running parameter~\cite{Babu:1993qv,Antusch:2001ck} and its one-loop RGE is approximately given by
\begin{equation}
\label{eq:kRG}
\beta_{\kappa}=\frac{1}{16\pi^{2}}\left(-3g^{2}+2\lambda_{H}+6y^{2}_{t}\right)\kappa.
\end{equation}
In the equation above, $g$ denotes the SU(2)$_{W}$ gauge coupling and $y_{t}$ the top Yukawa coupling. This is the only new parameter added to the SM and its value is very small since it has to yield the active neutrino masses. Its  contribution to the running of $\lambda_H$ is of order $v^{2}\kappa^2$ and thus negligible. The coupling $\kappa$ increases with energy due to $y_t$ being large albeit slowly. Moreover, the rest of the SM couplings run undisturbed. However, above the scale $M$ the neutrino Yukawa couplings start to run and affect the running of  $\lambda_H$ much like the top quark Yukawa coupling. Hence, the stability of the electroweak vacuum will set a limit on how large $Y_\nu$ can be if we demand that the presence of neutrinos does not make the electroweak vacuum unstable too soon. This was first studied in \cite{CCIQ} and later in~\cite{Rodejohann:2012px} with an emphasis on the Dirac Yukawa matrix dependence. Furthermore, the work in~\cite{Elias-Miro} covers in depth the case where the scalar potential is metastable with a lifetime longer than the age of the universe and in the presence of three degenerate right-handed neutrinos. In this work, we will extend their studies using the full 2-loop RG running of the SM couplings and ${\mathbf{Y}}_\nu$ and absolute stability in order to contrast with the next scenario. We also implement realistic neutrino mass matrices that encode the data from neutrino oscillations experiments~\cite{Bertuzzo:2013ew}. For simplicity we take ${\mathcal{M}}_M$ to be proportional to the unit matrix with scale given by $M_{N}$. To be consistent we require that $M_{N}<\Lambda_{I}$ where the electroweak vacuum becomes unstable.

An important quantity in our analysis is $\Tr [\kappa(m_h)]=\frac{2}{v^2}\sum_{i}^3 m_i$ where $m_i$ denotes the active neutrino masses. An upper limit on the above sum constrained by cosmology and astrophysics is given in \cite{cosmo}. We are mindful that the above is not without theoretical assumptions, i.e. the validity of the $\Lambda_{CDM}$ cosmology model. For degenerate neutrino masses we have $m_{\mathrm{cosmo}}=\sum_i m_i= 3m + \frac{\delta m^2}{2m} +\frac{\Delta m^2}{2m}$
where $m$ is the common mass and $\delta m^2$ and $\Delta m^2$ are the solar and atmospheric neutrino oscillation frequencies respectively. The most recent values for these parameters can be found in \cite{ospar}. For a normal hierarchy $m_{\mathrm{cosmo}}\simeq m_0\left( 1 +2\sqrt{1-\frac{\Delta m^2}{m_0^{2}}}\right)$ where $m_0$ is the heaviest of the three neutrinos. For the inverted hierarchy $m_{\mathrm{cosmo}}\simeq m_0\left( 2+\sqrt{1-\frac{\Delta m^2}{m_0^{2}}} \right)$. In both cases the smaller $\delta m^2$ term is dropped.
 Furthermore, we use the neutrino mass matrix elements found in~\cite{Bertuzzo:2013ew} and define at the Majorana mass scale $y^{2}_{\ell \ell}=2\frac{M_{N}}{v^{2}}m_{\ell \ell}$ where
 $\ell=e,\mu,\tau$. In addition, our study is carried out using the RGEs outlined in~\cite{Pirogov:1998tj} and where to one-loop order, the running of $\lambda_{H}$, $\Tr \left[{\mathbf{Y}}_\nu^\dagger {\mathbf{Y}}_\nu\right]$ and the diagonal elements of the neutrino Yukawa coupling matrix are approximately given by
\begin{eqnarray}
\frac{d \lambda_{H}}{dt}&\approx&\frac{1}{(4\pi)^{2}}\left[\frac{9}{4}\left(\frac{3}{25}g^{4}_{1}+\frac{2}{5}g^{2}_{1}g^{2}_{2}+g^{4}_{2}\right)-\left(\frac{9}{5}g^{2}_{1}+9g^{2}_{2}\right)\lambda_{H}+12y^{2}_{t}\lambda_{H}-12y^{4}_{t}+12\lambda_{H}^{2}\right. \nonumber \\
&+& \left. \lambda_{H}\Tr \left[{\mathbf{Y}}_\nu^\dagger {\mathbf{Y}}_\nu\right]\cdot\theta_{M_{N}}-4\left(y^{4}_{ee}+y^{4}_{\mu\mu}+y^{4}_{\tau\tau}\right)\cdot\theta_{M_{N}}\right] \nonumber \\
\frac{dy_{ii}}{dt}&\approx&\theta_{M_{N}}\cdot\frac{y_{ii}}{(4\pi)^{2}}\left[\frac{3}{2}y^{2}_{ii}+\left(3y^{2}_{t}+ \Tr \left[{\mathbf{Y}}_\nu^\dagger {\mathbf{Y}}_\nu\right]\right)-\left(\frac{9}{20}g^{2}_{1}+\frac{9}{4}g^{2}_{2}\right)\right]\nonumber \\
\frac{d \Tr \left[{\mathbf{Y}}_\nu^\dagger {\mathbf{Y}}_\nu\right]}{dt}&\approx&\theta_{M_{N}}\cdot\frac{2}{(4\pi)^{2}}\left[\left(3y^{2}_{t}+\Tr \left[{\mathbf{Y}}_\nu^\dagger {\mathbf{Y}}_\nu\right]\right)\Tr \left[{\mathbf{Y}}_\nu^\dagger {\mathbf{Y}}_\nu\right]+\left(\frac{9}{20}g^{2}_{1}+\frac{9}{4}g^{2}_{2}\right)\Tr \left[{\mathbf{Y}}_\nu^\dagger {\mathbf{Y}}_\nu\right]\right. \nonumber \\
&+&\left.\frac{3}{2}\left(y^{4}_{ee}+y^{4}_{\mu\mu}+y^{4}_{\tau\tau}\right)\right],\label{eq:Eq3}
\end{eqnarray}
where $i=e,\mu,\tau$  and $\theta_{M_{N}}\equiv\theta(\mu-M_{N})$ is a step function accounting for the threshold, $M_{N}$, at which the new neutrino Yukawa couplings begin to run. We have ignored the contributions from the bottom quark and $\tau$ lepton Yukawa couplings.
\begin{figure}[ht]\centering
\includegraphics[width=0.5\textwidth]{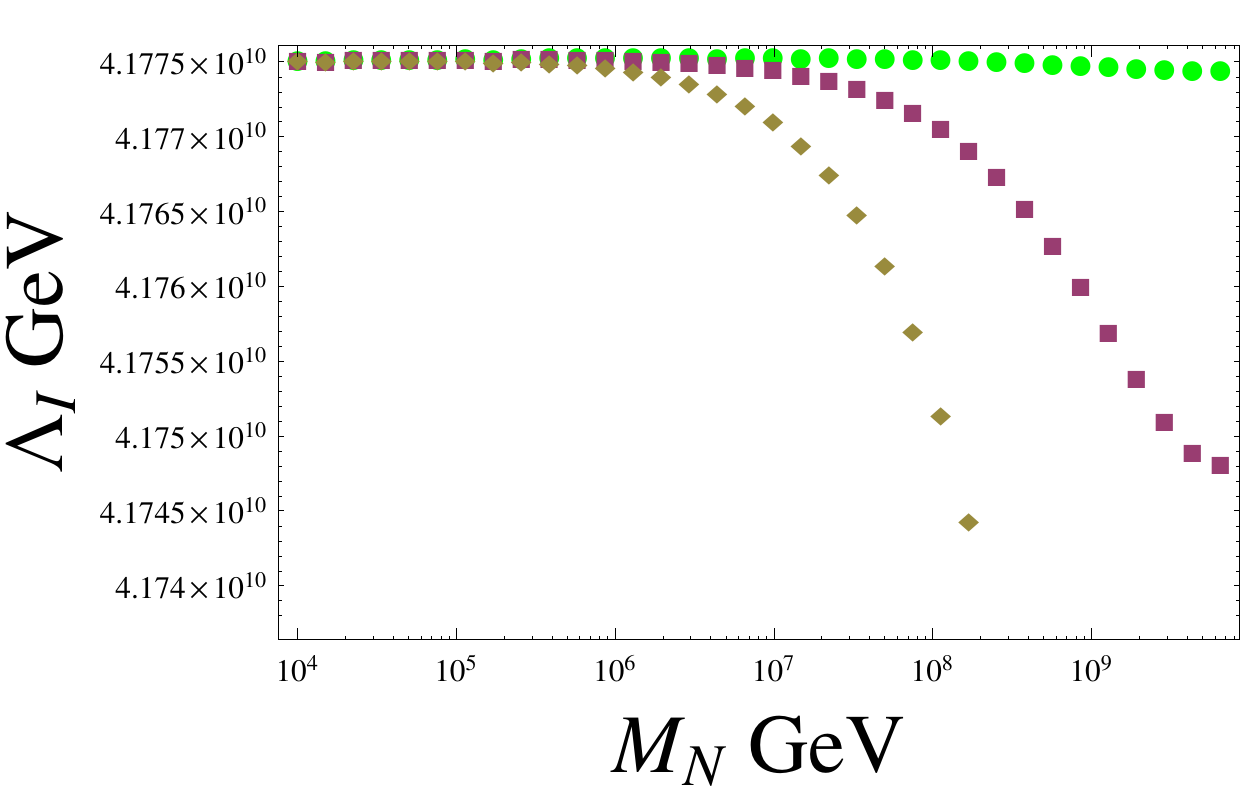}~~
\includegraphics[width=0.5\textwidth]{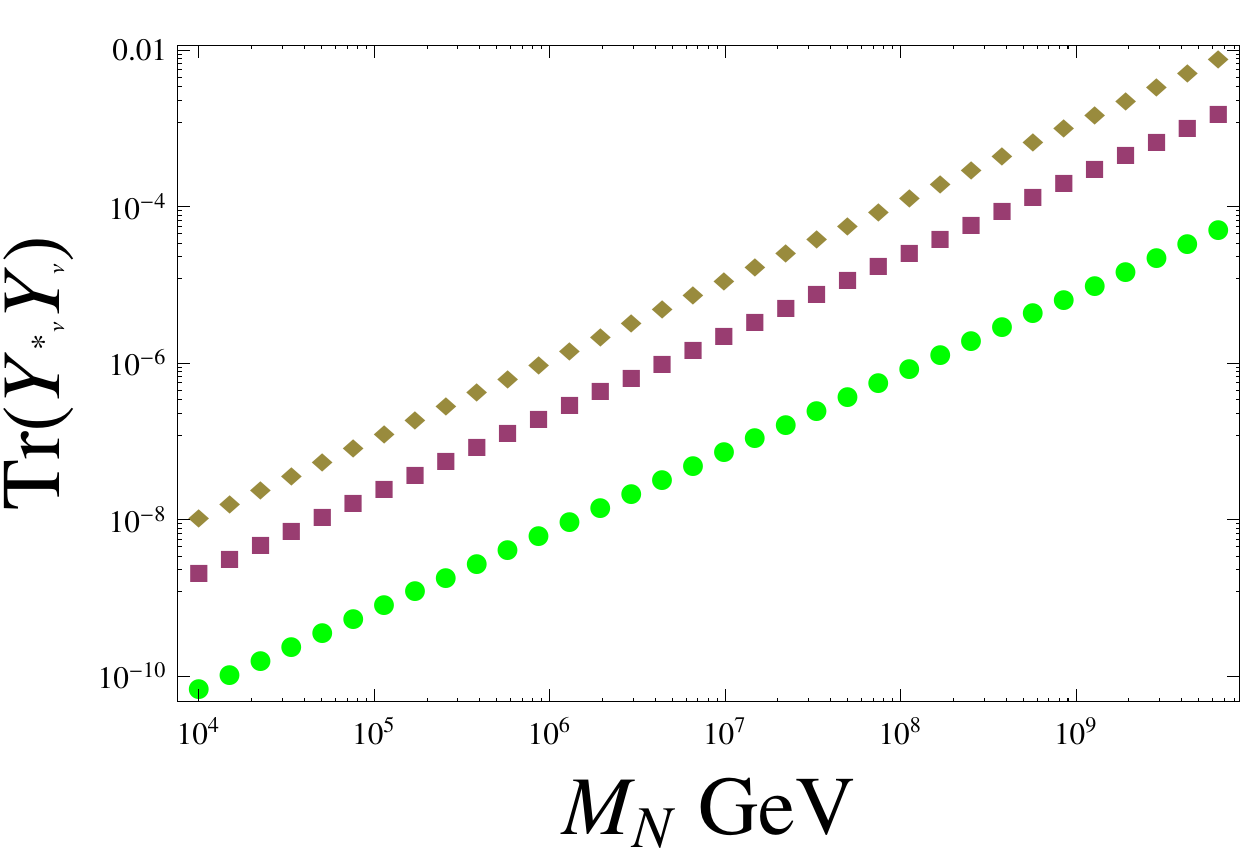}
\caption{\small  On the left (a) we plot the scale of instability, $\Lambda_{I}$ as a function of the bare mass$M_{N}$ using values for the sum of the active neutrino masses of $0.23$ eV (green), $2$ eV (red), and $10$ eV (brown). On the right (b) we show the value of $\Tr {\mathbf{Y}}_\nu^\dagger {\mathbf{Y}}_\nu$ as a function of $M_{N}$.  }\label{fig:Fig1}
\end{figure}

In Figure~\ref{fig:Fig1}(a) we show the value of $\Lambda_{I}$ as a function of $M_{N}$ using a value of the strong coupling constant $\alpha_{s}=0.1186\pm0.0006$~\cite{deBlas:2014mba} and a top quark mass of $173.21$ GeV~\cite{Agashe:2014kda}. The green circles correspond to a light neutrino mass consistent with the cosmological upper bound of $\sum_{i} m_{i}<0.23$ eV~\cite{Ade:2015xua}, while the red squares and brown diamonds correspond to the values of $m_0(m)=2,10$ eV respectively. The latter value is ruled out by current experimental bounds from tritium decays \cite{tritium} and we intend it to be for illustration purposes only. For a $0.05-2$ eV neutrino, the Majorana mass scale is bounded by $10^{10}$ GeV. Beyond this scale, the value of $\lambda_{H}$ at the Majorana mass scale is negative. This places an upper bound on the value of $\Tr {\mathbf{Y}}_\nu^\dagger {\mathbf{Y}}_\nu$ at the Majorana mass scale of $\sim10^{-4}$ for $M_{N}\sim10^{10}$ GeV and $m_0 (m)=0.07$ eV. We show this in Figure~\ref{fig:Fig1}(b). Our results is in general agreement with that of \cite{Elias-Miro}. However, we used realistic neutrino mass matrices and did not assume degenerate active neutrinos.

Beyond a Majorana mass scale of $\sim 10^{10}$ GeV the value of $\lambda_{H}\left(M_{N}\right)$ is negative and high scale seesaw models break down if the scalar potential were to
remain stable as we require. We also see that the absolute value of $\mathbf{Y}_\nu$ is of order 0.01. This is three orders of magnitude lower than the perturbative bound. This is valuable information since we know of no direct measurement of ${\mathbf{Y}}_\nu$ that can be made due to the very massive nature of the $N_R$'s. This is a well known problem of the high scale seesaw model. Yet, neutrino Yukawa couplings are vital for Type I seesaw models.  In attempts to circumvent this, more elaborate schemes such as low scale seesaw models~\cite{Ibarra:2010xw,Chakrabortty:2010rq,Wei:2010ww,deAlmeida:2010qb,Zhang:2009ac,Dev:2009aw,Xing:2009hx}, inverse seesaw models~\cite{Mohapatra:1986aw,Mohapatra:1986bd} and also the left-right symmetric models~\cite{LRM} have been introduced. The latter gives rise to signatures with same sign leptons plus jets that can be searched for at LHC.  A detailed study using a simplified model approach was recently given in \cite{NPP} where many references can be found. As for the high scale case, which has the virtue of being simple, one can only rely on theoretical studies. Indeed we conclude that for the current preferred value of $\alpha_s$, electroweak stability would lead to seesaw scales approximately six orders of magnitude lower than the Grand Unified theory scale with neutrino Yukawa couplings of order ${\cal O}\left(10^{-2}\right)$. Recently, the authors in~\cite{Rose:2015fua} have analyzed the stability of the electroweak vacuum in the presence of a low-scale seesaw model. They find that low scale seesaw models are viable and do not disrupt the stability of the Higgs potential for $\Tr {[\mathbf{Y}}_\nu^\dagger {\mathbf{Y}}_\nu] < 0.4$. Their results are complementary to our findings.

\section{Stability in the Type I Seesaw with a Complex Electroweak Singlet Scalar}
 In Eq.~(\ref{eq:Lnu}) the heavy right-handed neutrino masses are introduced by hand  and represents the case of explicit lepton number breaking. However, these masses can also be generated by spontaneous symmetry breaking of lepton number. The simplest model that achieves this involves adding a complex SM gauge singlet scalar field, $S$. It has a Yukawa coupling to right-handed neutrinos given by $Y_N\ovl{N_R}N_R^c S$ and as such preserves a global $U(1)_{L}$ symmetry responsible for lepton number if $S$ has a global lepton number charge of two units. Upon breaking this symmetry a Goldstone boson, the majoron, will emerge \cite{CMP} which can serve as a candidate for dark radiation \cite{CNW,CN}. This scalar couples to the Higgs field via the term $\lambda_{HS} S^\dagger S H^\dagger H$. One thus expects the running of the SM couplings and the stability of the scalar potential to be different. In the following we will address these issues and present as a detailed RG analysis of this model.

The embedding of the majoron model into a more complete model is not the purpose of this work. Indeed the $U(1)_L$ can be replaced by any $U(1)_X$.
A well known example is a gauged $X=B-L$ symmetry where the singlet scalar can serve as the inflaton. Here we focus on the effects of a complex scalar gauge singlet on the electroweak vacuum stability captured by the simple majoron model and compare the results with the explicit mass case studied above. The majoron plays no role in this investigation.

The addition of a complex scalar $S$, singlet under the SM gauge symmetries and charged under a global $U(1)_{L}$ lepton number symmetry, can be parametrized by a scalar potential given by
\begin{equation}
V_{H,S}=\frac{\lambda_{H}}{2}\left(H^{\dagger}H-v^{2}/2\right)^{2}+\frac{\lambda_{S}}{2}\left(S^{\dagger}S-w^{2}/2\right)^{2}+\lambda_{HS}\left(H^{\dagger}H-v^{2}/2\right)\left(S^{\dagger}S-w^{2}/2\right).\label{eq:Eq4}
\end{equation}
Within this framework, for $\lambda_{H},\lambda_{S}>0$ and $\lambda_{HS}<\lambda_{H}\lambda_{S}$, the minimum of the potential is at $\left<H\right>=v/\sqrt{2}$ and $\left<S\right>=w/\sqrt{2}$. The mixing between the two neutral CP even scalars, is given by
\begin{equation}
\tan\theta=\frac{\lambda_{HS}vw}{\lambda_{S}w^{2}-\lambda_{H}v^{2}}.\label{eq:Eq5}
\end{equation}
In this work, we are interested in the limit where the lightest scalar is the SM-like Higgs boson. This limit is characterized by a large singlet vacuum expectation value, $w>>v$, and masses for the two CP even scalars given by
\begin{equation}
m^{2}_{h^{0}}\simeq\lambda_{H}v^{2},~~~~~~m^{2}_{S}\simeq\lambda_{S}w^{2}.
\end{equation}
This also implies small mixing of the SM-like Higgs with the heavy scalar and hence only small corrections to the Higgs couplings to other SM particles.
However, the presence of a heavy scalar, coupling at tree-level to the SM-like Higgs boson, may lead to a large positive contribution to the RGE  for $\lambda_{H}$~\cite{Gonderinger:2009jp,Lebedev:2011aq,Kadastik:2011aa,Gonderinger:2012rd,Chen:2012faa,Costa:2014qga,Costa:2015sva,Falkowski:2015iwa} as well as a tree-level threshold effect~\cite{EliasMiro:2012ay} that arises from the matching effect of the singlet at the energy $Q\sim m_S$ . Both are known to affect the stability of the Higgs potential. The authors in~\cite{EliasMiro:2012ay} have studied the effects from a threshold corrections to the Higgs quartic coupling, $\delta\lambda=\lambda^{2}_{HS}/\lambda_{S}$ against those that arise from positive contributions in the running of $\lambda_{H}$. They studied two interesting cases: One where $\lambda_{HS}>0$, where large threshold corrections increase the scale at which the Higgs potential develops an instability and the second where $\lambda_{HS}<0$. In the latter, the tree-level threshold effect is not sufficient to increase the instability scale and RG effects become important since the new stability condition, $\lambda_{H}>\delta\lambda$ must be satisfied at all scales.

We wish to study the effects on the instability scale of the SM scalar potential in the presence of a new heavy scalar, as the one discussed above, and also incorporate three Majorana right-handed neutrinos. Within this framework, a Majorana mass is dynamically generated once the new scalar develops a $vev$ with the coupling $Y_{N}\ovl{N_R}N_R^c S$. For simplicity we will take $Y_N$ to be real. The RG evolution of $\lambda_{H}$ will be modified at the scales $M_{N}$ and $m_{S}$ due to the presence of new fermion and scalar degrees of freedom, the former which will tend to destabilize the scalar potential. We analyze both the case where $\lambda_{HS}>0$ and $\lambda_{HS}<0$ and compare our results to a model with only a scalar gauge singlet.

The scale of instability can be calculated using the RG equations for the scalar and new Yukawa couplings which at one-loop order are given by
\begin{eqnarray}
\frac{d \lambda_{H}}{dt}&\approx&\frac{1}{(4\pi)^{2}}\left[\frac{9}{4}\left(\frac{3}{25}g^{4}_{1}+\frac{2}{5}g^{2}_{1}g^{2}_{2}+g^{4}_{2}\right)-\left(\frac{9}{5}g^{2}_{1}+9g^{2}_{2}\right)\lambda_{H}+12y^{2}_{t}\lambda_{H}-12y^{4}_{t}+12\lambda_{H}^{2}\right. \nonumber \\
&+&\left.\left( 2\lambda^{2}_{HS}+ \lambda_{H}\Tr \left[{\mathbf{Y}}_\nu^\dagger {\mathbf{Y}}_\nu\right]-4(y^{4}_{ee}+y^{4}_{\mu\mu}+y^{4}_{\tau\tau})\right)\cdot\theta_{M_{high}}\right] \nonumber \\
\frac{d\lambda_{HS}}{dt}&\approx&\theta_{M_{high}}\cdot\frac{1}{(4\pi)^{2}}\left[\frac{1}{2}\left(12y^{2}_{t}-\frac{9}{5}g^{2}_{1}-9g^{2}_{2}\right)\lambda_{HS}+4\lambda_{HS}\left(\frac{3}{2}\lambda_{H}+\lambda_{S}\right)+4\lambda^{2}_{HS}+4(Y^{2}_{1}+Y^{2}_{2}+Y^{2}_{3})\lambda_{HS}\right] \nonumber \\
\frac{d\lambda_{S}}{dt}&\approx&\theta_{M_{high}}\cdot\frac{1}{(4\pi)^{2}}\left[4\lambda^{2}_{HS}+10\lambda^{2}_{S}-2(Y^{4}_{1}+Y^{4}_{2}+Y^{4}_{3})+4(Y^{2}_{1}+Y^{2}_{2}+Y^{2}_{3})\lambda^{2}_{S}\right] \nonumber \\
\frac{dY_{i}}{dt}&\approx&\theta_{M_{high}}\cdot\frac{Y_{i}}{(4\pi)^{2}}\left[4Y^{2}_{i}+2(Y^{2}_{1}+Y^{2}_{2}+Y^{2}_{3})\right],\label{eq:Eq7}
\end{eqnarray}
\begin{figure}[ht]\centering
\includegraphics[width=0.50\textwidth]{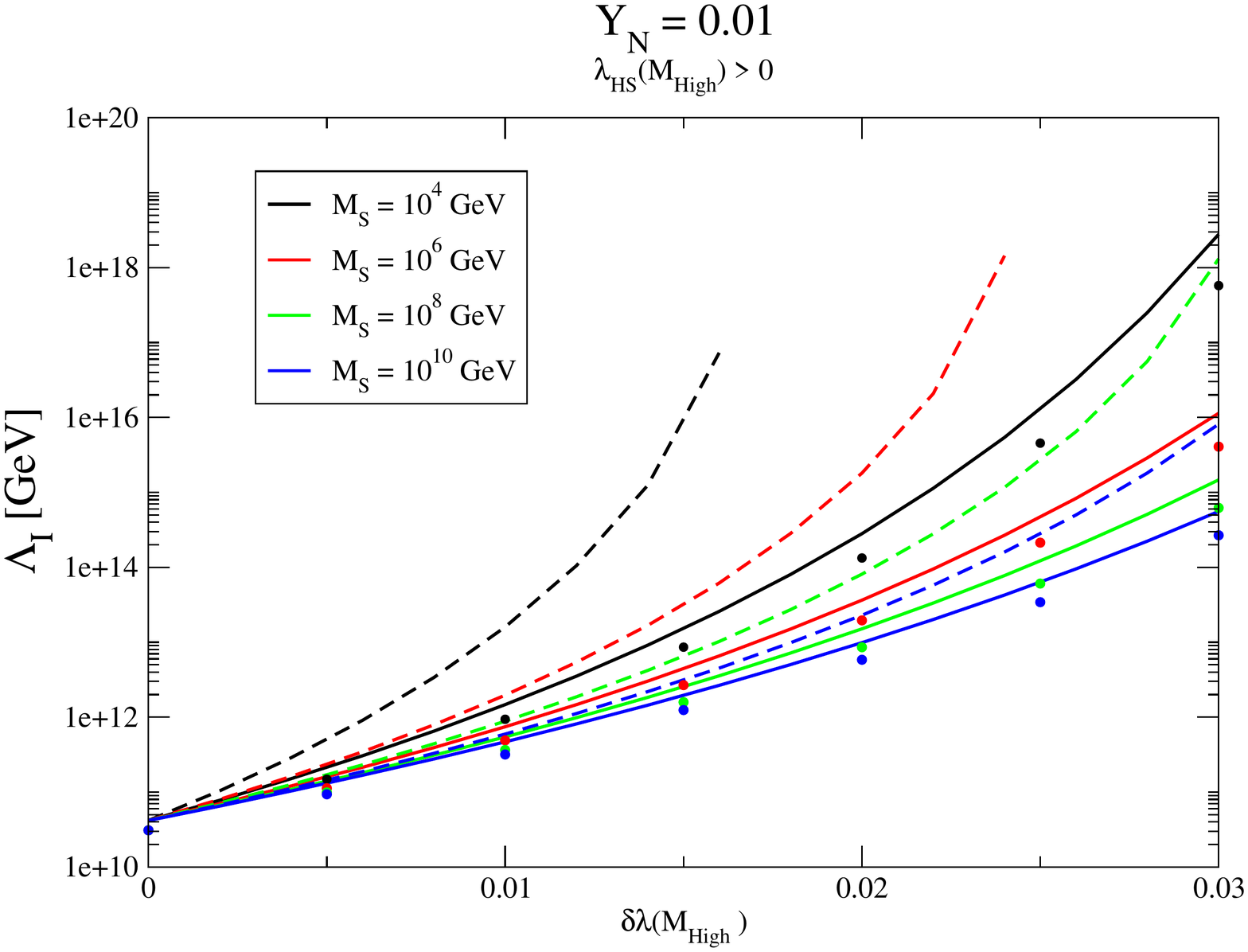}

\includegraphics[width=0.48\textwidth]{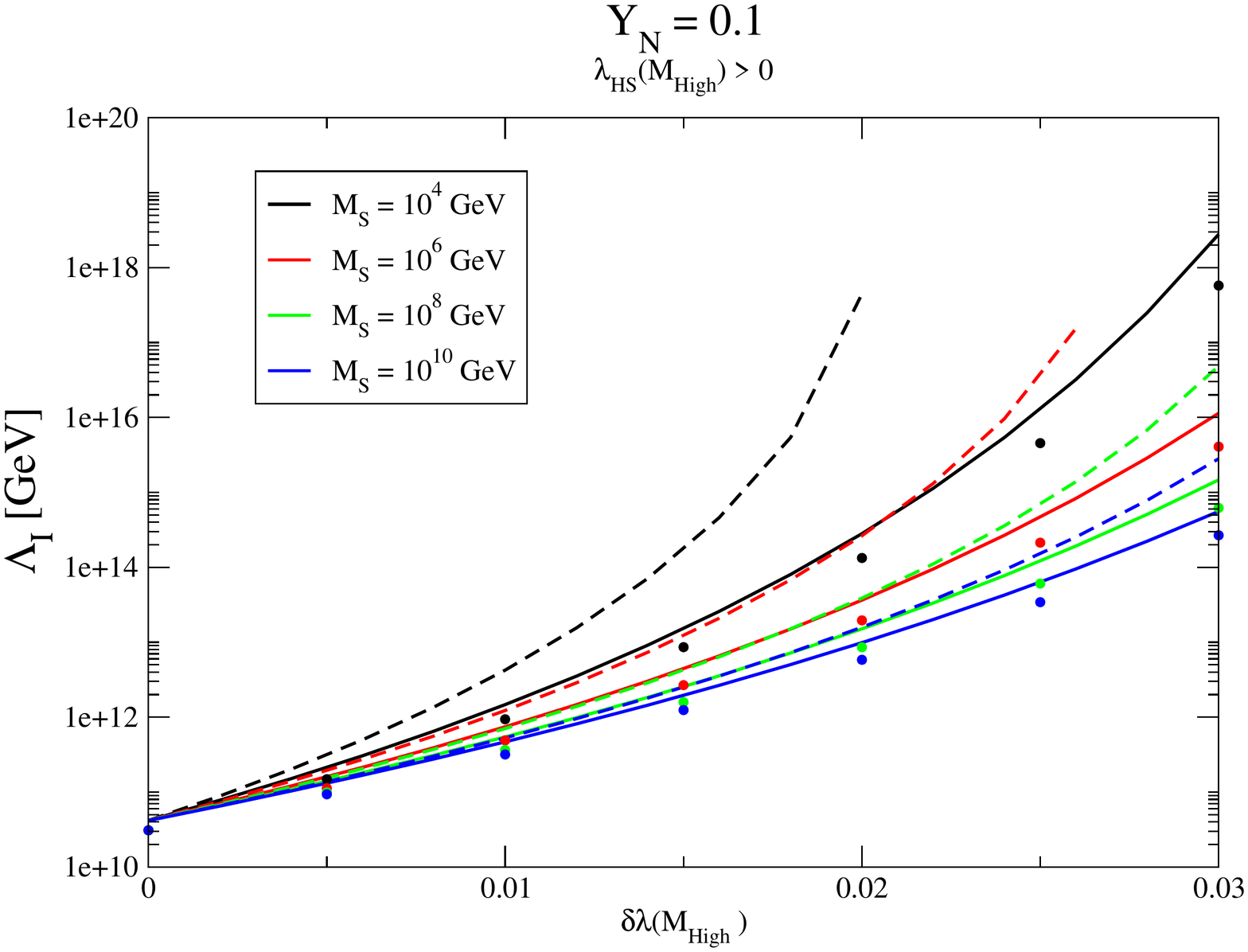} ~~
\includegraphics[width=0.48\textwidth]{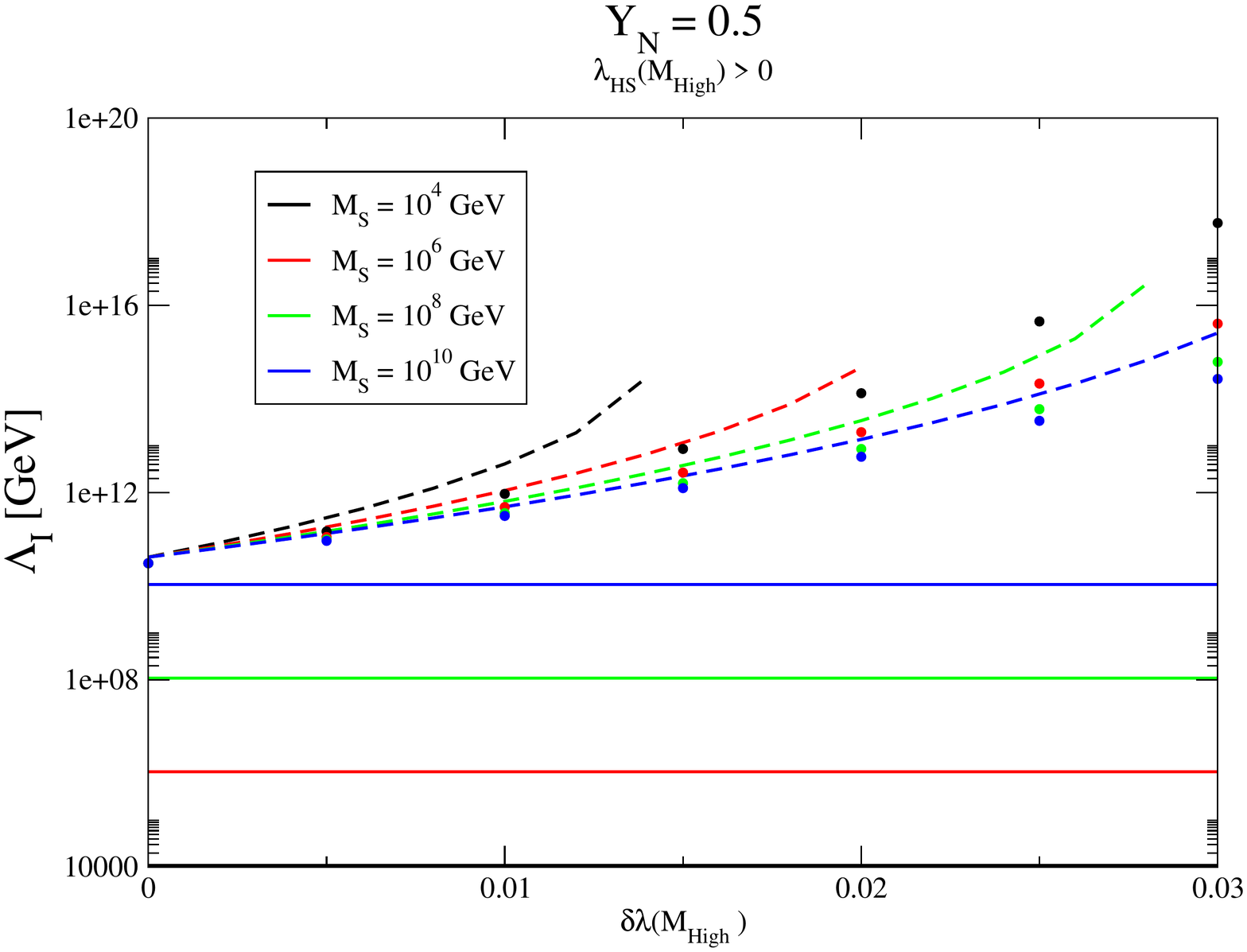} ~~
\caption{\small The scale of instability, $\Lambda_{I}$ within Type I seesaw models extended by a complex scalar gauge singlet as function of the threshold contribution to $\lambda_{H}$, $\delta\lambda$ defined at $M_{high}$ for different values of $m_{S}$. The figure on the top panel, (a), corresponds to $\lambda_{HS}>0$ and $Y_{N}=0.01$ while the bottom panel, (b-c) correspond to $Y_{N}=0.1,0.5$ respectively.   }\label{fig:Fig2}
\end{figure}
where $\theta_{M_{high}}$, is a step function accounting for the threshold, $M_{high}=Min\{M_{N},m_{s}\}$, at which the new Yukawa coupling the singlet to the right-handed Majorana neutrinos and scalar potential parameters begin to run. We assume that $S$ couples diagonally to all three right-handed neutrinos with universal strength $Y_{i}=Y_{N}$ for all $i$, and the $vev$ of $S$ sets the Majorana mass scale given by $M_{N_{i}}=2Y_{N}\left<S\right>=(2/\sqrt{2})Y_{N}w$. The running of the Yukawa couplings between left- and right-handed neutrinos to the SM-like Higgs, $Y_{\nu}$, are given as Eq.~(\ref{eq:Eq3}) with $M_{N}\to M_{high}$. The instability scale for $\lambda_{HS}>0$ is defined as the energy scale where either $\lambda_{H}$ or $\lambda_{S}$ first vanish, rendering the scalar potential unstable and invalidating the seesaw mechanism in models were one demands absolute stability of the scalar potential.  For $\lambda_{HS}<0$, the condition $\lambda_{H}>\delta\lambda$ must also be met for all scales together with $\lambda_{S}>0$. By inspection of Eq.~(\ref{eq:Eq7}) we expect that for large values of $Y_{N}\left(M_{high}\right)$, $\lambda_{S}$ will be driven towards negative values very rapidly given that it runs with the fourth power of $Y_{N}$. In fact, for $\lambda_{HS}>0$, three effects define the scale where the scalar potential becomes unstable:
 \begin{itemize}
 \item A potentially large tree-level threshold effect to the SM Higgs quartic coupling. This tends to increase the instability scale of the SM Higgs.
 \item A large value of $\lambda_{S}\left(M_{high}\right)$ which provides a large positive contribution to the running of $\lambda_{H}$ which also improves the stability of the SM Higgs.
 \item  A large negative contribution to the running of $\lambda_{S}$ from large values of $Y_{N}\left(M_{high}\right)$. This tends to drive $\lambda_{S}$ towards zero at scales near large $M_{high}$.
 \end{itemize}

There is a second effect that disfavors large values of $Y_N$. The last equation in Eq.(\ref{eq:Eq7}) show that these couplings hit a Landau pole very quickly. If this happens within the stability region of the scalar potential, the theory would become strongly coupling. Although a very interesting scenario, it is beyond the scope of this work. For small values of $Y_N$, $\lambda_S$ can also arrive at the Landau pole too soon. These conspire to limit the range of interesting values of $Y_N$ and $\lambda_S$.

\begin{figure}[ht]\centering
\includegraphics[width=0.50\textwidth]{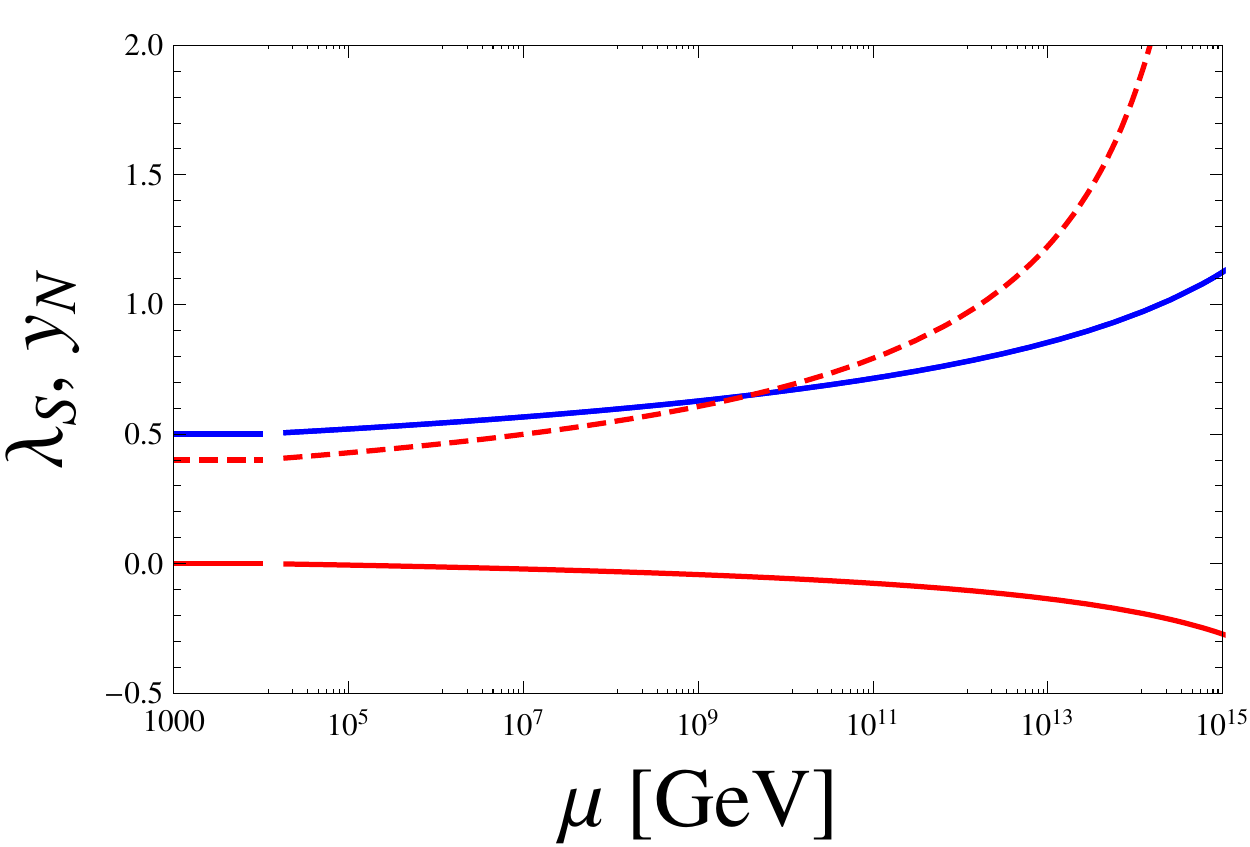}
\caption{\small Running of $Y_{N}$ with $Y\left(M_{high}\right)=0.5$ denoted by the blue solid line while in red we show the running of $\lambda_{S}$ with $\lambda_{S}(M_{high})=10^{-4}$ (solid) $\lambda_{S}(M_{high})=0.4$ (dashed). }\label{fig:Fig3}
\end{figure}

  A qualitatively picture of these effects for $\lambda_{HS}>0$ is depicted in Figures~\ref{fig:Fig2}(a)-(c) where we plot the scale of instability, $\Lambda_{I}$, as a function of $\delta\lambda\left(M_{high}\right)$ for values of $Y_{N}\left(M_{high}\right)=0.01,0.1,0.5$ respectively. In the figures, the dashed lines correspond to $\lambda_{S}\left(M_{high}\right)=0.4$, while the solid lines correspond to $\lambda_{S}\left(M_{high}\right)=0.0001$. The solid dots correspond to the complex scalar extension of the SM parametrized by Eq.~(\ref{eq:Eq4}); i.e. without the Majorana neutrino effects. The finite nature of the various curves denote the following: If $\delta\lambda$ increased either we obtain stability up to the Planck scale with couplings either perturbative or non-perturbative at the Planck scale, or one of the quartics goes to zero and then hits a Landau Pole driving the other couplings towards a Landau Pole. Basically, a larger $\delta\lambda$ leads to a larger $\lambda_{HS}$ which also affects the running. From the figures it is clear that a large $\lambda_{S}\left(M_{high}\right)$ aids a lot at stabilizing the Higgs potential for small values of $Y_{N}\left(M_{high}\right)$. This is due to the positive contribution to the running of $\lambda_{H}$ which together with the tree-level threshold effect help increase the stability of the scalar potential compared to $\Lambda^{\text{SM}}_{I}\approx10^{10}$ GeV. However, as $Y_{N}\left(M_{high}\right)$ increases, the effect from the tree-level threshold effect shift remains fairly constant since $\lambda_{S}$ goes to zero near $M_{high}$ for small $\lambda_{S}\left(M_{high}\right)$. For larger $\lambda_{S}\left(M_{high}\right)$, the contributions from $\lambda_{S}$ to the running of $\lambda_{S}$ for a large range of energies and it is eventually overtaken by the contribution from $Y_{N}$. This aids at stabilizing the scalar potential at high energies. We can observe this effect in Figure~\ref{fig:Fig3} for $m_{S}=10^{4}$ GeV. In the figure, we denote with the solid blue line the running of $Y_{N}$ with $Y_{N}\left(M_{high}\right)=0.5$ and in solid and dashed red the running of $\lambda_{S}$ for $\lambda_{S}\left(M_{high}\right)=10^{-4},0.4$ respectively. The stabilization of the scalar potential is helped by the large value of $\lambda_{S}$ at $M_{high}$ and this effect becomes more relevant for higher $M_{high}$ since the new couplings run for a smaller range up to the Planck scale.

\begin{figure}[ht]\centering
\includegraphics[width=0.50\textwidth]{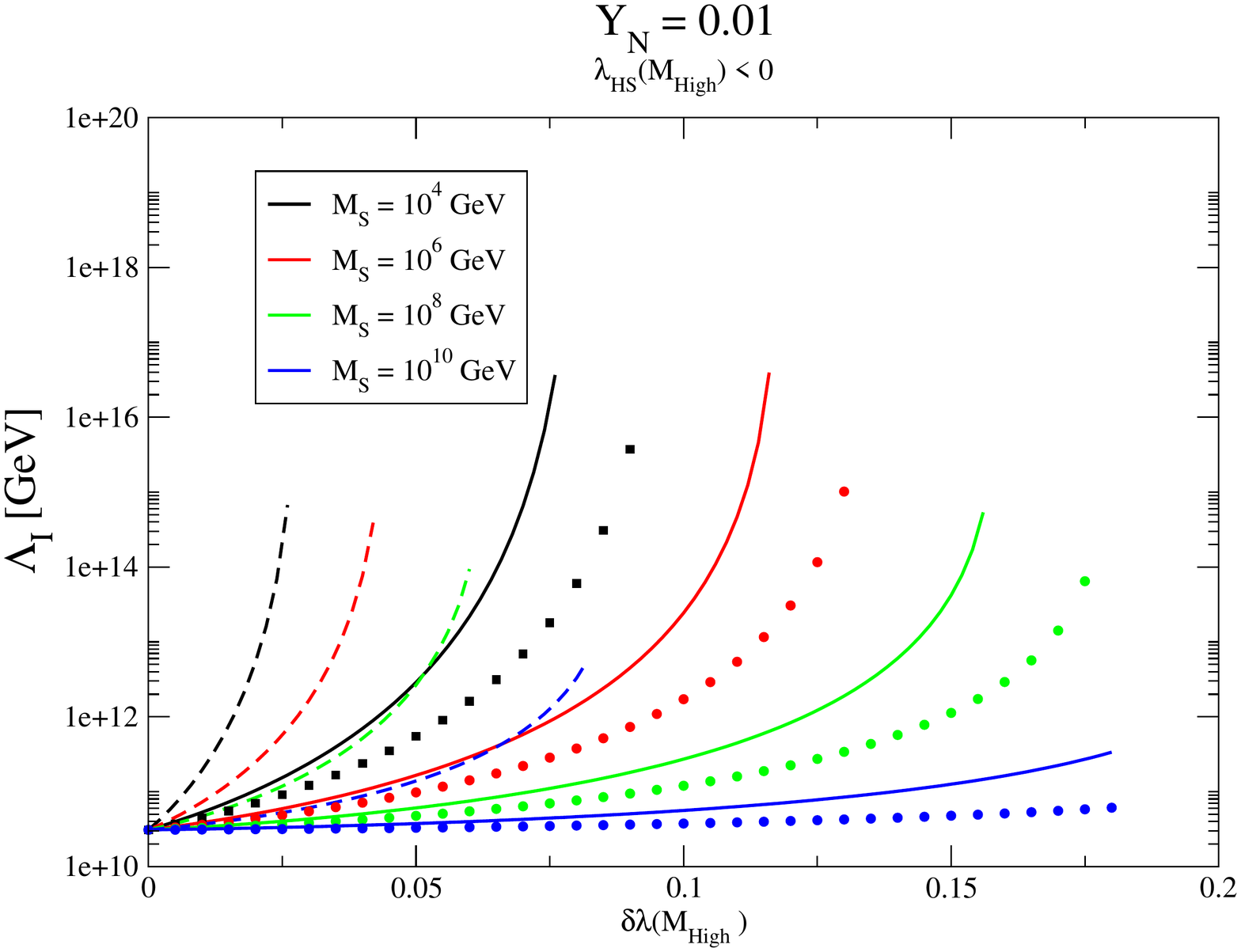}

\includegraphics[width=0.48\textwidth]{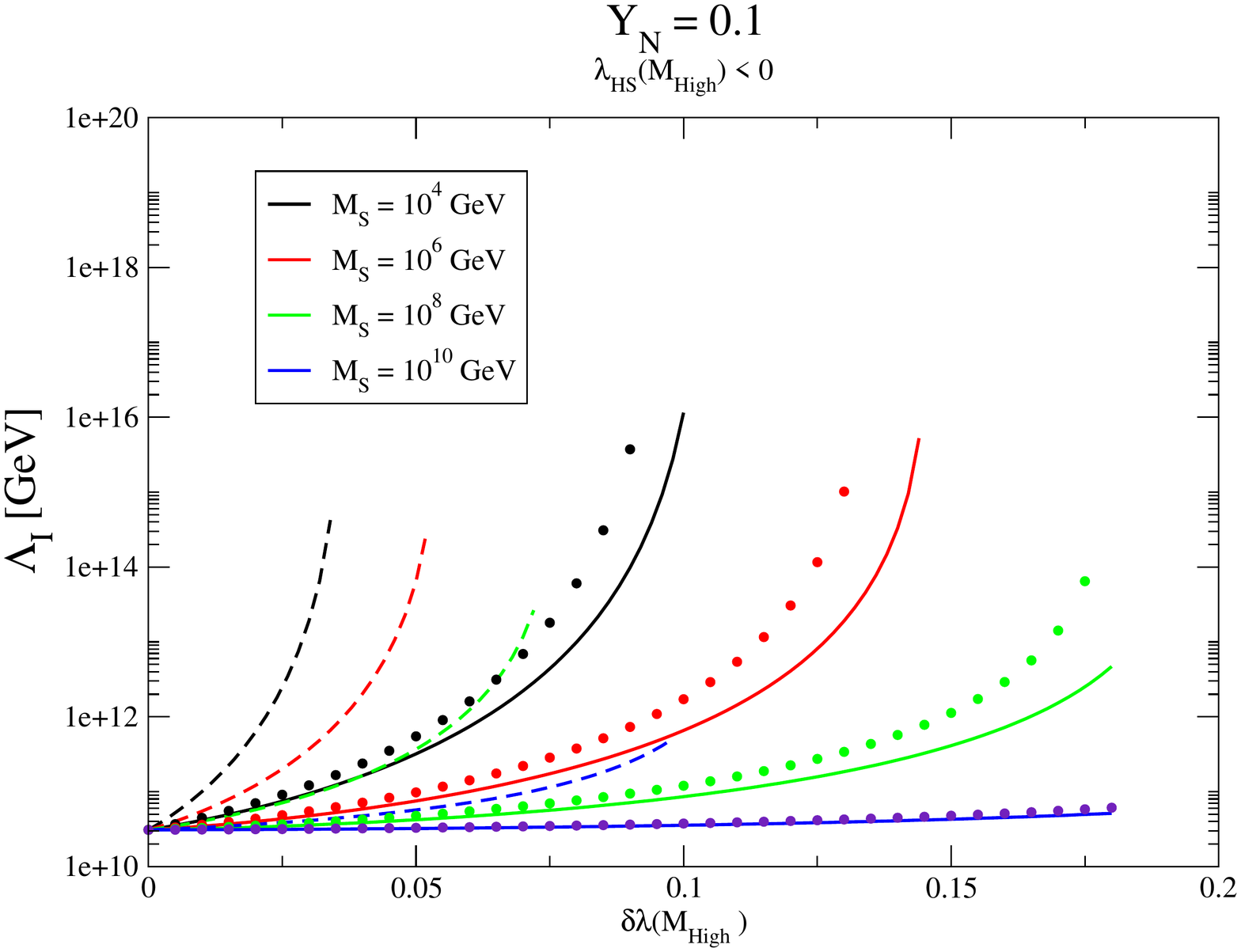} ~~
\includegraphics[width=0.48\textwidth]{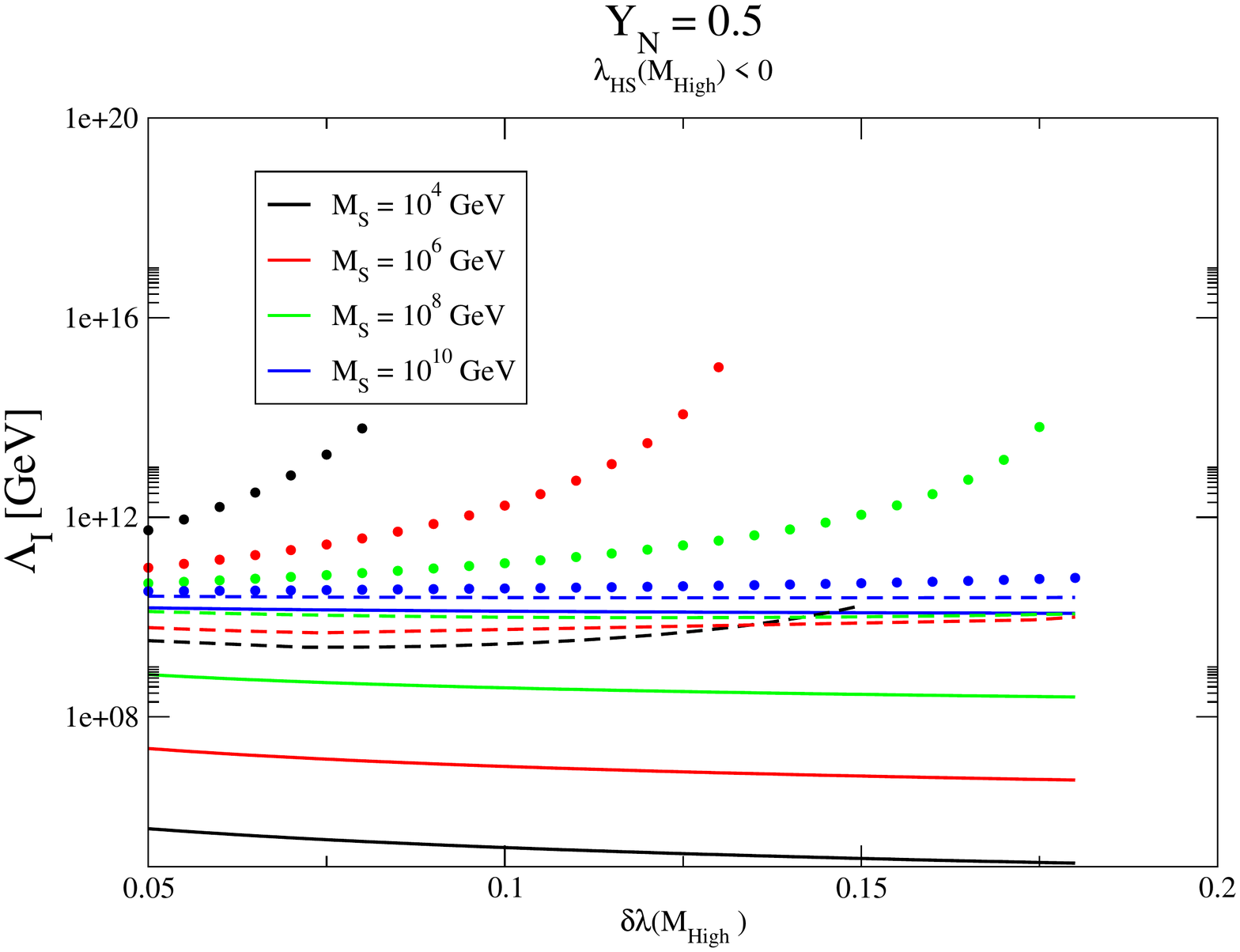} ~~
\caption{\small  The scale of instability, $\Lambda_{I}$ within Type I seesaw models extended by a complex scalar gauge singlet as function of the threshold contribution to $\lambda_{H}$, $\delta\lambda$ defined at $M_{high}$ for different values of $m_{S}$. The figure on the top panel, (a), corresponds to $\lambda_{HS}<0$ and $Y_{N}=0.01$ while the bottom panel, (b-c) correspond to $Y_{N}=0.1,0.5$ respectively.}\label{fig:Fig4}
\end{figure}

For $\lambda_{HS}<0$, the new stability condition, $\lambda_{H}>\lambda^{2}_{HS}/\lambda_{S}$ together with $\lambda_{S}>0$ must be satisfied. The new condition on $\lambda_{H}$ has the effect of destabilizing the scalar potential at lower energies for small values of $\lambda_{S}\left(M_{high}\right)$ since a positive contribution to $\lambda_{H}$ from $\lambda_{S}$ does not help to drive $\lambda_{H}$ above $\delta\lambda$ at scales above $M_{high}$. For large values of $\lambda_{S}\left(M_{high}\right)$ an additional positive contribution to the running of $\lambda_{H}$ raises the scale of instability for smaller values of $\delta\lambda\left(M_{high}\right)$. In Figures~\ref{fig:Fig4}(a)-(c) we plot the scale of instability, $\Lambda_{I}$, as a function of $\delta\lambda\left(M_{high}\right)$ for values of $Y_{N}\left(M_{high}\right)=0.01,0.1,0.5$ respectively. In the figures, the dashed lines correspond to $\lambda_{S}\left(M_{high}\right)=0.4$ while the solid lines correspond to $\lambda_{S}\left(M_{high}\right)=0.02$. The solid dots correspond to the complex scalar extension of the SM parametrized by Eq.~(\ref{eq:Eq4}).

\begin{figure}[ht]\centering
\includegraphics[width=0.40\textwidth]{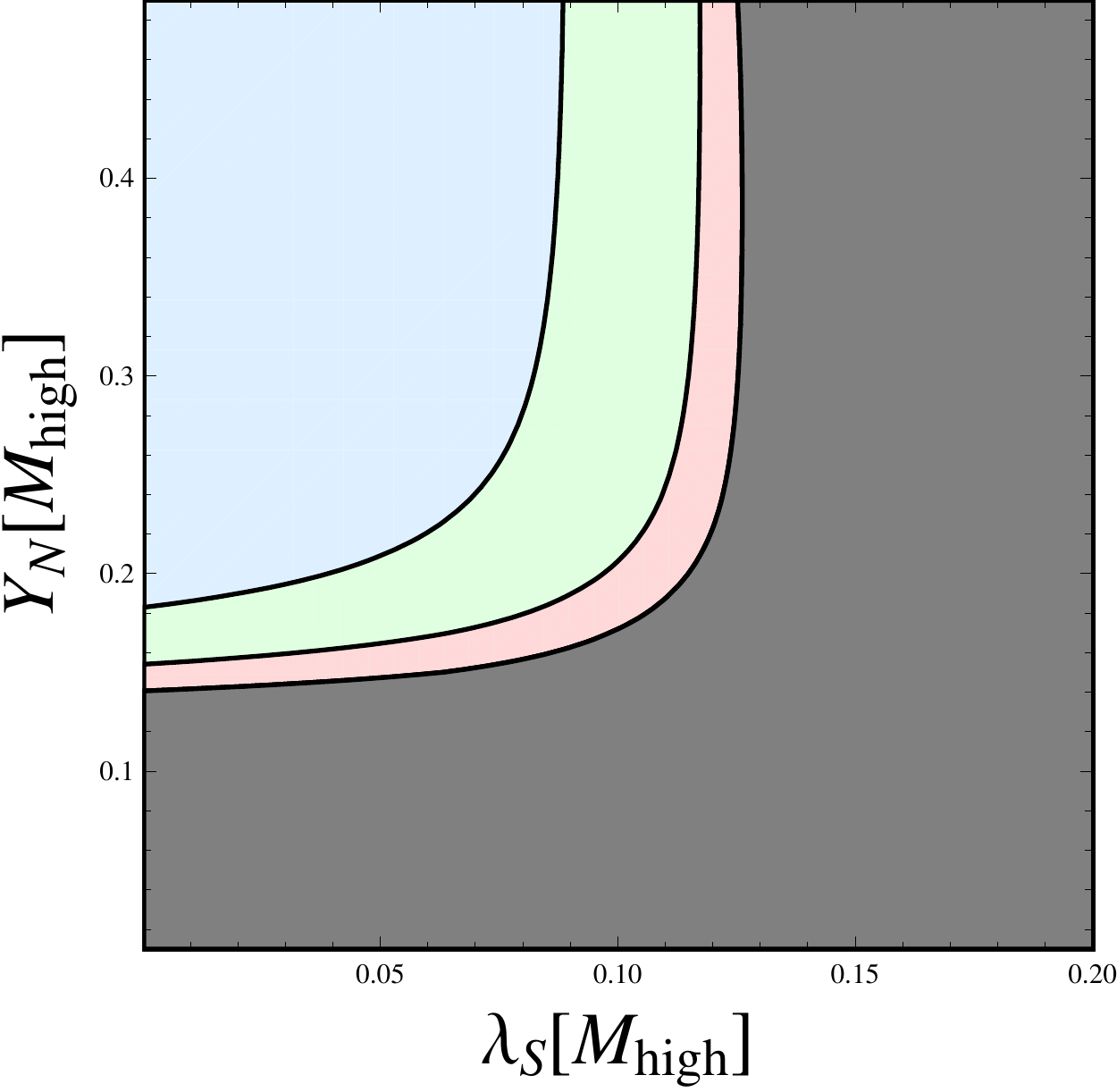}
\caption{\small  Regions of parameter space where $\Lambda_{I}\gtrsim10^{10}$ GeV for $m_{S}=10^{4}$ GeV shown in gray. The regions in light red, green and blue are stacked being the gray region and correspond to $m_{S}=10^{6},10^{8},10^{10}$ GeV respectively with the light blue region extending over the entire grid}\label{fig:Fig5}
\end{figure}

We note that our results suggest that $\lambda_{S}$ cannot be too small unless $Y_{N}$ is simultaneously small since otherwise $\lambda_{S}$ will be driven negative very quickly. To see this we scan over $\lambda_{S}\left(M_{high}\right)$ and $Y_{N}\left(M_{high}\right)$ and calculate the scale of instability. Our results are shown in Figure~\ref{fig:Fig5} for $\delta\lambda=0.015$ and $\lambda_{HS}>0$. The gray region corresponds to regions of parameter space where $\Lambda_{I}\gtrsim10^{10}$ GeV for $m_{S}=10^{4}$ GeV. The regions in light red, green and blue are stacked behind the gray region and correspond to $m_{S}=10^{6},10^{8},10^{10}$ GeV respectively with the light blue region extending over the entire grid.

\begin{figure}[ht]
\begin{center}
\resizebox{\columnwidth}{!}{\input{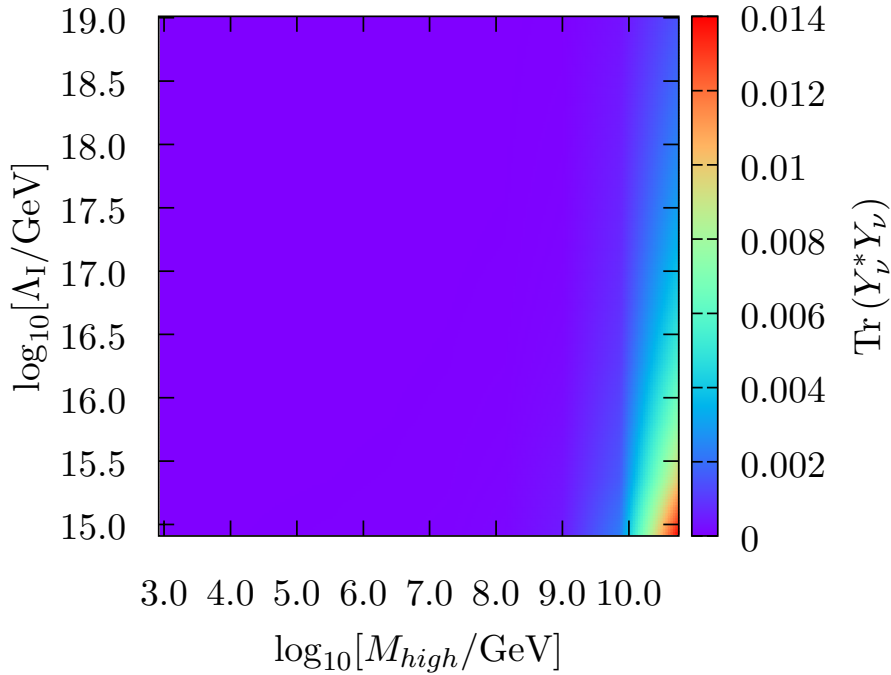}}
\end{center}
\caption{\small The value of $\Tr[\mathbf{Y}^\dagger_{\nu}\mathbf{Y}_\nu]\left(M_{High}\right)$ in the $\log[\Lambda_{I}/~\text{GeV}]-\log[M_{high}/~\text{GeV}]$ plane. In this figure we choose $\lambda_{HS}>0$, $\delta\lambda=0.030$, $\lambda_{S}=0.4$ and $Y_{N}=0.1$ defined at the scale $M_{high}$.}
\label{fig:Fig6}
\end{figure}

Next we turn our attention to $\mathbf {Y}_\nu$. In the above we have seen that spontaneously generating heavy Majorana neutrino masses can raise the instability scale of the electroweak vacuum with the help of the added scalars; especially for $\lambda_{HS}>0$. One can then increase the upper bound on $\Tr [{\mathbf{Y}_\nu^\dagger}{\mathbf{Y}_{\nu}}]$ compared to the case of explicit mass terms. This can be seen in Figure \ref{fig:Fig6} with $Y_{N}=0.1$.
Indeed the bound is $\Tr[{\mathbf{Y}_\nu^\dagger}{\mathbf{Y}_{\nu}}]\lesssim0.01$, higher than before but is only for a relatively large $M_{high}$. In this figure, only points which lead to an instability scale below or at the Planck scale with perturbative couplings up to the Planck scale are shown. Regions of parameter space for which couplings are not perturbative below the Planck scale are not shown. These points tend to be for low $M_{high}$. For very large $M_{high}$, the couplings run as in the SM and the instability scale is below $M_{high}$. This region of parameter space corresponds to points where $M_{high}\gtrsim10^{10}$ GeV, and are not accepted either. The window of allowed values of $M_{high}$ gets smaller for larger values of $\lambda_{S}$.
\begin{figure}
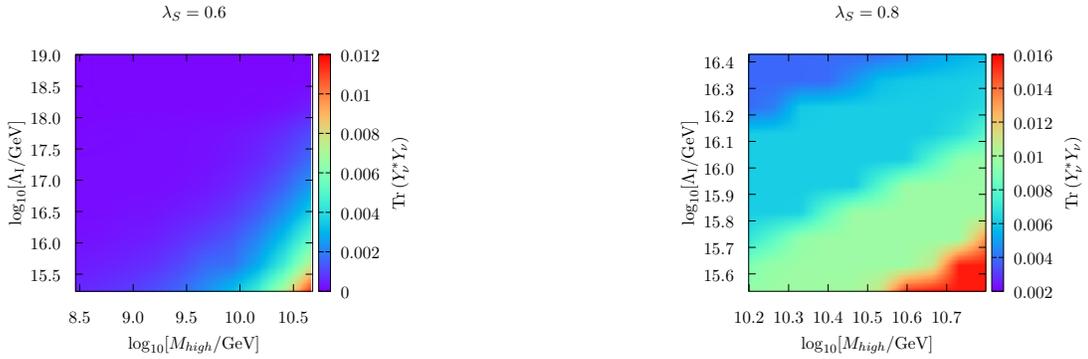

\centering
\begin{minipage}[t]{7.5cm}
\resizebox{\columnwidth}{!}{\input{yukawa_singlet_lamS_0_6}}
\end{minipage}
\hfill
\begin{minipage}[t]{7.5cm}
\resizebox{\columnwidth}{!}{\input{yukawa_singlet_lamS_0_8}}
\end{minipage}
\caption{\small Same as in Figure~\ref{fig:Fig6} with $\lambda_{S}=0.6$ (left (a)) and $\lambda_{S}=0.8$ (right (b)).}
\label{fig:Fig7}
\end{figure}
 This can be seen in Figures~\ref{fig:Fig7}(a) and (b) with $\lambda_{S}=0.6$ and $0.8$ respectively. Values of $\lambda_{S}$ beyond $\sim0.9$ do not lead to a viable seesaw. Furthermore, increasing $Y_{N}$ only decreases the upper bound on $\lambda_{S}$.

\section{Discussion}
We conclude that in the two heavy Majorana mass generation mechanisms we have studied, the neutrino Yukawa couplings will remain perturbative. We find that for high-scale Type I seesaw models in which lepton number is explicitly broken by Majorana bare mass terms the stability is lowered due to the Dirac Yukawa couplings, $\mathbf {Y}_\nu$. We also find that $\Tr [\mathbf{Y}^\dagger_{\nu}  \mathbf{Y}_\nu]\lesssim10^{-3}$ in order for the SM electroweak vacuum stability not be worsened. In the scenario where right-handed neutrinos masses are generated spontaneously through a singlet scalar $vev$, the electroweak vacuum can be stable up to the Planck scale through threshold effects from decoupling a heavy scalar and positive contributions to the running of $\lambda_{H}$ which ultimately depends on the scalar mixing parameter $\lambda_{HS}$. We find that the scale at which the scalar potential becomes unstable is not improved over $\Lambda^{\text{SM}}_{I}$ if the quartic coupling of the singlet is small while $Y_{N}$ is large. This result is worsened for light singlet scalar masses; since $\lambda_{S}$ will be driven negative at very low energy scales. We also set the limit $\Tr [\mathbf{Y}^\dagger_{\nu}\mathbf{Y}_\nu]\lesssim0.01$. This is because both $Y_N$ and $\lambda_S$ must not be too large as not to hit
the Landau pole within stability region. Thus, the viable boundary values of $M_N$ are low and consequently small values of $\Tr [\mathbf{Y}^\dagger_{\nu}\mathbf{Y}_\nu]$ from $\kappa$.


\section*{Acknowledgements}

This work is supported in parts by the National Science and Engineering Council of Canada.

\end{document}